\documentclass[prl,twocolumn,superscriptaddress,showpacs,floatfix]{revtex4-1}
\usepackage{xcolor}
\usepackage{graphics}
\usepackage{graphicx}
\usepackage{amsmath}
\usepackage{amsthm}
\usepackage{amssymb}
\usepackage{braket}

\newcommand{\tpsi}{\tilde{\psi}}
\newcommand{\tphi}{\tilde{\phi}}

\newcommand{\tb}{\textcolor{blue}}
\newcommand{\tg}{\textcolor{violet}}
\usepackage[utf8]{inputenc}
\usepackage[english]{babel}
\newtheorem*{theorem}{Theorem}

\begin{document}

\title{\textcolor{black}{Closed form Eigenvalues of Randomly Segmented Tridiagonal quasi-Toeplitz Matrices: Random Rouse block copolymer}}
\author{S.S. Ashwin}
\email{ss.ashwin@gmail.com}
\affiliation{Theoretical Biophysics Lab,\\
Department of Applied Physics, Nagoya University,
Nagoya, Japan}
\affiliation{Center for Computational Natural Sciences and Bioinformatics\\
IIIT Hyderabad, Gachibowli, Hyderabad}


\date{\today}

\begin{abstract}
	\textcolor{black}{We calculate the eigenvalues of a class of random matrices, namely the randomly segmented tridiagonal quasi-Toeplitz (rstq-T) matrix, in exact closed-form. The contexts under which these matrices arise are ubiquitous in physics. In our case, they arise when studying the dynamics of a Rouse polymer embedded in random environments. Unlike in the case of Rouse polymers in homogeneous environments, where the dynamics give rise to a circulant matrix and the diagonalization is achieved easily via a Fourier transform, analytical diagonalization of the rstq-T matrix has remained unsolved thus far. We analytically calculate the spectral distribution of the rstq-T matrix, which is able to capture the effect of  disorder on the modes.} 
\end{abstract}
\maketitle
\textcolor{black}{Exact results to elucidate mechanisms in the physics of disordered systems~\cite{Gautam,Bovier} are limited. Random matrix theory (RMT)~\cite{MLMehta,Vivo,Potters} provides a powerful framework for analyzing the statistical properties of complex systems in general and disordered systems in particular (~\cite{PJForrester2003,PJForrester1} and references therein).  Wigner~\cite{Wigner1955} was the first to introduce Random matrices (RM) in order to understand the spectra of heavy nuclei. The complexity of the Hamiltonian of heavy nuclei and its diagonalization was a formidable task in the 1950s. By modeling the Hamiltonian using a Hermitian matrix with random elements, Wigner was able to explain several features of the nuclear spectra. Dyson built on this work and developed the idea of ensembles of unitary matrices and an associated unique measure~\cite{PJForrester1,Dyson1962}. RMT has since been applied in diverse fields such as neural networks~\cite{Aljadeff,Coolen,Kuczala,Rajan,Louart}, quantum systems~\cite{Sachdev}, spin glasses~\cite{Mezard,Bray,Kosterlitz}, ecology~\cite{RMMay,Barabas}, number theory~\cite{Sarnak1996,Keating1995,Katz1999}, and chaotic systems~\cite{Bohigas}.}

\textcolor{black}{ 
The eigenspectra for several classes of RM are known~\cite{MLMehta, PJForresterBook}. The techniques to calculate RM spectra can be broadly categorized as ``histograms without histogramming"~\cite{Potters,Albrecht}, as they evaluate spectra without explicit knowledge of the eigenvalues. 
Recently, closed-form expressions for the the largest eigenvalue of RM with generalized correlations has been found~\cite{Baron2022}.
Closed-form expressions for eigenvalues offer more direct insight into the physical parameters that control the spectra. Closed-form expressions are  particularly useful when dominant eigenvalues, smallest eigenvalues, singularities or eigenvectors~\cite{Tao} are of interest. No exact closed-form eigenvalues are known for any class of RM under general conditions, to the best of the authors' knowledge. Although random tridiagonal matrices, the simplest non-trivial RM class, have been intensively studied since the 1960s~\cite{Dean, Hori, Matsuda} following Dyson's classical works~\cite{Dyson}, no explicit expression for their eigenvalues has been found, even though their spectra have been significantly understood.}

\textcolor{black}{
	Another class of matrices that have found wide applications in physics are the Toeplitz matrices~\cite{Toep1911}. The eigenvalues of Toeplitz matrices~\cite{Gray} have exact and elegant closed-form expressions, and are ubiquitous in physics. Perhaps the most known appearance of Toeplitz matrices is in the celebrated Kaufman and Onsager correlation function for the Ising model~\cite{Onsager}. Toeplitz matrices are diagonal-constant matrices, while random Toeplitz matrices are matrices where the constant diagonals are chosen from a distribution. These matrices have only recently been numerically investigated~\cite{EBogomolny}. Tridiagonal Toeplitz matrices arise in uniform systems with one-dimensional topology~\cite{Hartwig, Qin, Ivanov, Gray, Keat}. The constant diagonals of Toeplitz matrices embed some symmetry of the systems they arise in. Breaking this symmetry at longer ranges gives rise to a form of disorder. For instance, in systems where local ordering and long-range disorder co-exist, and the underlying topology is one-dimensional, such as 1D tight-binding models with disorder~\cite{Dunlop} and 1D wave propagation in layered random media, one would expect to come across a class of matrices that are diagonal constant up to a range but taken from a distribution. This would result in the {\it randomly segmented tridiagonal quasi-Toeplitz} matrices (rstq-T). In this letter, we provide an exact closed-form expression for the eigenvalues of the rstq-T class of RM for a given realization of randomness and analytically deduce the spectra for the random distribution. We show the application of these results to the dynamics of polymers in heterogeneous media, which is central to many aspects of soft-matter physics and chromatin dynamics, in particular.}

\textcolor{black}{Chromatin packing directly affects epigenetics~\cite{EpigenRev, Virk2020}, so it's crucial to understand. Studying how chromatin moves can help us understand how it is packed in live cells. Single-particle tracking of nucleosomes has greately enhanced our understanding of how chromatin moves~\cite{AshPNAS, AshRev, AmitaiRev, Itoh2021, Maeshima2020}. By adding fluorescent tags to protein octamers (nucleosomes) that chromatin is wrapped around, live cell tracking of the nucleosomes becomes feasible. The cellular nuclear environment is composed of phase-separated liquid droplets~\cite{Courchaine}. The motion of nucleosomes embedded in these diverse liquid droplets are a possible source of multiple diffusivity peaks~\cite{AshPNAS, AshRev}. Single-particle tracking reveals that nucleosome motion is subdiffusive and its mean squared displacement (MSD) behaves as $\sim t^{0.5}$, with time ($t$)~\cite{Maeshima2019, AshPNAS, MaeshimaII2019, Sasai2016}. The {\it Rouse bead spring polymer} model (RBP)\cite{Verdier1966,Doi} has the same MSD exponent as nucleosomes, making it a good minimal model for nucleosome dynamics. RBP and its variations have been studied to understand chromatin conformation capture data\cite{HolcmanPRE} and folding and fluctuations in yeast chromatin~\cite{Socol}. RBP assumes a constant damping coefficient, and its dynamical equations have a circulant matrix form, which is diagonalizable using a Fourier transformation, resulting in uncoupled Langevin equations with exact solutions. Assuming dissolved chromatin loops\cite{Nozaki2017} and neglecting ATP activity~\cite{Winkler,Rabin}, a minimal polymer model for nucleosome dynamics in the heterogeneous nuclear environment is the {\it random Rouse block copolymer} (rRBC), here the segments of the polymer are embedded in different liquid droplets. The dynamical equations matrix for rRBC yields the rstq-T matrix, which we analytically diagonalize.}

The bead-spring model assumes over-damped Brownian dynamics, where nearest neighboring beads are connected by springs with strength $\kappa$. Non-neighboring beads don't interact energetically or volumetrically.. The polymer consists of $N_{\tb{m}}$ beads divided into $m$ {\it bead-segments}, each characterized by the number of beads $n_{\tb{i}}$ and damping coefficient $\xi_{\tb{i}}$, $\forall i\in\{1,\dots,m\}$, with $\sum_{i=1}^{m}n_{\tb{i}}=N_{\tb{m}}$. A map $f:l\rightarrow i$ is defined between the bead index $l\in\{1,\dots,N_{\tb{m}}\}$ and the bead-segment index $i\in \{1,\ldots,m\}$ it belongs to. The bead positions are denoted by ${\bf R}=[\vec{R}_{\tb{1}},\dots,\vec{R}_{\tb{N}_{\tg{m}}}]^{T}$. The Langevin equations are:

\begin{equation}
	  \label{LE} 
	  \frac{d {\bf R}}{dt}=-\kappa{\bf \Xi^{-1}ZR} + {\bf \Xi^{-1} F}
  \end{equation}
Here ${\bf \Xi} =diag[\xi_{\tb{1}},...\xi_{\tb{1}},\xi_{\tb{2}},..,\xi_{\tb{2}},...\xi_{\tb{m}}]$ is a $N_{\tb{m}}\times N_{\tb{m}}$ diagonal matrix and ${\bf Z}$ is the bead connectivity matrix with periodic boundary conditions.
\begin{align}
	  {\bf Z} &= \begin{bmatrix}
		  -2& 1 & 0 &..  .&..1 \\
		  1 & -2 & 1& .. &0\\
           \vdots \\
		  1&0 &..&  ..1&-2
         \end{bmatrix}
  \end{align}

  \textcolor{black}{Any bead $l$ is subject to $\delta$-correlated white noise $\eta_{\tb{l}}(t)$ with $<\eta_{\tb{l}}(t)>=0$. For any other bead $j$, $<\eta_{\tb{l}}(t)\eta_{\tb{j}}(t')>=2k_BT\delta_{\tb{lj}}\delta(t-t')$ ~\cite{Doi}. The term $[\Xi^{-1}{\bf F}]_{\tb{l}}=\eta_{\tb{l}}(t)\sqrt{ k_BT}/\xi_{\tb{f(l)}}$, where $k_B$ is the Boltzmann constant. We define:}
  \begin{equation}
	  {\bf M} = -\kappa{\bf \Xi^{-1}Z }
 \end{equation}	  
Let {\bf S} be the similar matrix which diagonalizes {\bf M}. \textcolor{black}{Application of this transform on (\ref{LE}) would result in an uncoupled set of} Langevin equations with exact solutions~\cite{Chaikin}
\begin{equation}
	\frac{d {\bf X}}{dt}=-\kappa{\bf S M S^{-1} X} + {\bf S \Xi^{-1} F}
	\label{singlemode}
\end{equation}	
Here ${\bf X}={\bf SR}=[\vec{X}_{\tb{1}},..\vec{X}_{\tb{N}_{\tg{m}}}]^{T} $. \textcolor{black}{The linear transform ${\bf S \Xi^{-1} F}$ leaves the nature of the noise invariant~\cite{Doi}}. The task now is to diagonalize {\bf M}. We do the diagonalization in the limit $N_{\tb{m}}>>1$, and the number of beads in each bead-segment $n_{\tb{i}}\geq 2$ $\forall i\in\{1,\dots,m\}$ need not be large.
\textcolor{black}{ The periodic unit entries in the first and last row of ${\bf Z}$ are ignored since the error in the eigenvalues is of order $O(1/N_{\tb{m}}^3)$~\cite{Grudsky}.} Then,
 \begin{equation}
 [{\bf M}]_{\tb{ij}} = \kappa\delta_{\tb{i-1,j}} \xi^{-1}_{\tb{f(j)}}+ \kappa\delta_{\tb{i+1,j}}\xi_{\tb{f(j)}}^{-1}-2\kappa\delta_{\tb{ij}}\xi_{\tb{f(j)}}^{-1}
 \end{equation}
  \textcolor{black}{$\forall i,j\in\{1,\ldots,N_{\tb{m}}\}$. With $\kappa=1$}. The submatrix ${\bf M}_{\tb{u:v}}$ includes all beads from bead index $u$ to $v>u$ in {\bf M} is :
\begin{equation}
	  {\bf M}_{\tb{u:v}} =\left[
              \begin{array}{c c c c c} 
		      2\xi_{\tb{f(u)}}^{-1}& -\xi_{\tb{f(u)}}^{-1} & 0 &.. 0 .&0 \\
		  -\xi_{\tb{f(u+1)}}^{-1} & 2\xi_{\tb{f(u+1)}}^{-1}  & -\xi_{\tb{f(u+1)}}^{-1}  .. \\
		  0&\ddots&\ddots&\ddots&\ddots\\  
		  0 &0 &\cdot& -\xi_{\tb{f(v)}}& 2\xi_{\tb{f(v)}}^{-1} 
               \end{array}
               \right]
\end{equation}

\textcolor{black}{If $\xi_{\tb{f(j)}}$ is the same for $u\leq j\leq v$, but different from $\xi_{\tb{f(u-1)}}$ and $\xi_{\tb{f(v+1)}}$, we call the submatrix ${\bf M}_{\tb{u:v}}$ a {\it segment}. If $\xi_{\tb{i}}$ were the same $\forall i\in\{1,\ldots,m\}$, then ${\bf M}$ would be a tridiagonal Toeplitz matrix with a single segment. However, if $\xi_{\tb{i}}$ are chosen from a random distribution, then  ${\bf M}$ becomes a {\it randomly segmented-quasi-uniform tridiagonal Toeplitz} matrix (rsqt-T).}
\textcolor{black}{We aim to first derive exact expressions for the eigenvalues of the matrix ${\bf M}$ with $m$ segments for a realization of $\xi_{\tb{i}}$ $\forall i\in\{1,\ldots,m\}$. The characteristic equation is $\psi_{1:\tb{N}_{\tg{m}}}=\det[({\bf M}_{\tb{1:N}_{\tg{m}}}-\lambda {\bf I})=0$. By studying the cases for $m=1$ and $m=2$, we establish a pattern that leads to a theorem for the $m^{th}$ case. This yields a sequence of random matrix products that, in the continuum limit, become an integral equation for the spectral distribution of the rsqt-T matrix.}

{\it {\bf Case $m=1$:}} Consider the characteristic equation for $ \psi_{\tb{1:n}_{\tg{1}}}(\lambda)=\det[({\bf M}_{\tb{1:n}_{\tg{1}}}-\lambda {\bf I})]$. For the first segment, $\xi_{\tb{f(j)}}=\xi_{\tb{1}}$ with $1\leq j\leq n_{\tb{1}}$. ${\bf M}_{\tb{1:n}_{\tg{1}}}$ is a Toeplitz matrix. Using the cofactor expansion on the last row of $({\bf M}_{\tb{1:n}_{\tg{1}}}-\lambda {\bf I})$ \textcolor{black}{in terms of $\tpsi=\frac{1}{\xi_{\tb{1}}^{n_{\tb{1}}}}\psi(\lambda)$}:  
    \begin{equation}
	    \tpsi_{\tb{1:n}_{\tg{1}}}(\lambda)=(2-\lambda\xi_{\tb{1}})\tpsi_{\tb{1:n}_{\tg{1}}-\tb{1}}(\lambda) - \tpsi_{\tb{1:n}_{\tg{1}}-\tb{2}}(\lambda)
    \end{equation}	
Using the Chebyshev recurrence relation of the second kind~\cite{Mason}, we can identify $2\cos(k)=2-\lambda\xi_{\tb{1}}$, $\tpsi_{\tb{1:0}}=1$, and $\tpsi_{\tb{1:1}}=(2-\lambda \xi_{\tb{1}})$. This yields the characteristic equation as the Chebyshev polynomial~\cite{Mason} of the second kind, given by $ \tpsi_{\tb{1:n}_{\tg{1}}} = \mathcal{U}_{\tb{n}_{\tg{1}}}(k)\equiv\frac{\sin[(n_{\tb{1}}+1)k]}{\sin(k)}$. The roots of $\tpsi_{\tb{1:n}_{\tg{1}}}=0$ correspond to the eigenvalues, $\lambda_{\tb{p}} = \frac{4}{\xi_{\tb{1}}}[\sin^2(k_{\tb{p}}/2)]$ where $k_{\tb{p}}=p\pi/(n_{\tb{1}}+1)$ $\forall p\in\{1,\dots,n_{\tb{1}}\}$.

{\it {\bf Case $m=2$:}} Consider the matrix below for the case $N_{\tb 2}=n_{\tb{1}}+n_{\tb{2}}$:
\begin{equation}
   {\bf M}_{\tb{1:n}_{\tg{1}}\tb{+n}_{\tg{2}}}=\left[
     \begin{array}{c|c}
      {\bf M}_{\tb{1:n}_{\tg{1}}} &  
     \begin{array}{c c c} 
	     0 & \cdots & 0\\ 
	     \vdots & \vdots & \vdots\\ 
     0 & \cdots & 0\\ 
     1/\xi_{\tb{1}} & 0 & 0 
  \end{array}  \\ 
  \hline 
  \begin{array}{c c c}
   0&\cdots & 1/\xi_{\tb{2}}\\
     0 & \cdots & 0\\ 
     \vdots & \vdots & \vdots\\ 
	  0 & \cdots & 0\\ 
   \end{array} & {\bf M}_{\tb{n}_{\tg{1}}\tb{+1:n}_{\tg{1}}+\tb{n}_{\tg{2}}} 
 \end{array} 
\right] 
\end{equation}
Using Laplace expansion in the $n_{\tb{1}}+1$ row and Leibniz formula for determinants~\cite{Kulkarni}. Redefining, $\tpsi_{n}=(\frac{1}{\xi_2})^n\det(M_{\tb{n}}-\lambda{\bf I})$ $\forall$ $n\in \{1,\dots,N_m\}$ we have:
 \begin{eqnarray}
	 \label{kulk}
	 \tpsi_{1:\tb{n}_{\tg{1}}\tb{+n}_{\tg{2}}}(\lambda) &=&  \tpsi_{\tb{1:n_{\tg{1}}}}(\lambda) \mathcal{U}_{\tb{n_{\tg{2}}}}(\lambda)\\\nonumber
				   &-& w_1 \tpsi_{\tb{1:n_{\tg{1}}-1}}(\lambda)\mathcal{U}_{\tb{n}_{\tg{2}}-\tb{1}}(\lambda)\\\nonumber
 \end{eqnarray}	

\textcolor{black}{We define, $w_i=\frac{\xi_{\tb{i+1}}}{\xi_{\tb{i}}} ~\forall i \in \{1,\ldots,m\}$. Since, $\tpsi_{\tb{1:n}_{\tg{1}}\tb{+n}_{\tg{2}}}$ can be expanded using $\mathcal{U}_{\tb{n_{\tg{1}}}},\mathcal{U}_{\tb{n_{\tg{2}}}}$ it is a function of cosines $2\cos(k)$. First, we need to find the $k_{\tb{p}}$ roots of $\tpsi_{\tb{1:n}_{\tg{1}} \tb{+n}_{\tg{2}}}=0$ $\forall p\in \{1,\ldots,n_{\tb{1}}+n_{\tb{2}}\}$, from which we can obtain the corresponding $\lambda$'s.}

\textcolor{black}{Banchi and Vaia ~\cite{Banchi} showed that eigenvalues of matrices with mostly uniform Toeplitz structure and a small non-uniform part can be expressed as Toeplitz eigenvalues that are phase-shifted. We adopt their method as our starting point since all segments in our case are Toeplitz}.
 
Consider the following identity between Chebyshev polynomials:
\begin{equation}
	\label{id1}
	\mathcal{U}_{\tb{n-1}}(k)=\cos(k) \mathcal{U}_{\tb{n}}(k)-\mathcal{T}_{\tb{n+1}}(k)
\end{equation}	
The Chebyshev polynomial of the first kind $\mathcal{T}_{\tb{n}}(x)=\cos(nx)$. We apply identity (\ref{id1}) to (\ref{kulk}) to obtain:
\begin{eqnarray}
	\tpsi_{\tb{1:n}_{\tg{1}}+\tb{n}_{\tg{2}}}&=&cos[(n_{\tb{2}}+1)k](w_{\tb{1}} \tpsi_{\tb{1:n}_{\tg{1}}\tb{-1}})\\\nonumber
			&+&\sin[(n_{\tb{2}}+1)k](\tpsi_{\tb{1:n}_{\tg{1}}}-w_{\tb{1}} \cos(k)\tpsi_{\tb{1:n}_{\tg{1}} \tb{-1}})\\\nonumber
			&=&\sin((n_{\tb{1}}+n_{\tb{2}}+1)k + \phi_{\tb{2}})/\sin(k)\nonumber
   \end{eqnarray}
where $\phi_{\tb{2}}$ is given by:
\begin{equation}
	\phi_{\tb{2}}=-n_{\tb{1}}k + \tan^{-1}\left(\frac{w_1 \tpsi_{\tb{1:n}_{\tg{1}}\tb{-1}}\sin(k)}{\tpsi_{\tb{1:n}_{\tg{1}}}-w_{\tb{1}} \cos(k)\tpsi_{\tb{1:n}_{\tg{1}}\tb{-1}}}\right)
\end{equation}
The roots of $\psi_{\tb{1:n}_{\tg{1}}+\tb{n}_{\tg{2}}}=0$ are given by $k_p=(p\pi-\phi_{\tb{2}})/(n_{\tb{1}}+n_{\tb{2}}+1)$, with $\forall p \in \{1,\dots,n_{\tb{1}}+n_{\tb{2}}\}$. The eigenvalues are obtained by identifying $2-\xi_{f(p)} \lambda=2\cos(k_{\tb{p}})$,
then $\lambda_{\tb{p}}=\frac{4}{\xi_{\tb{f(p)}}}\sin^2(k_{\tb{p}}/2)$.

{\it {\bf Case m}:} The emerging pattern from $m=1$ and $m=2$ suggests a generalization for any $m$. In order to do this we introduce two functions: 
\begin{eqnarray}
	F_{\tb{n}_{\tg{i}}}(k) &=& \mathcal{U}_{\tb{n}_{\tg{i}}} -w_{\tb{i} } \cos(k)\mathcal{U}_{\tb{n}_{\tg{i}}-\tb{1}}\\\nonumber
	Q_{\tb{n}_{\tg{i}}}(k) &=&  \mathcal{T}_{\tb{n}_{\tg{i}}+\tb{1}} -w_{\tb{i}} \cos(k)\mathcal{T}_{\tb{n}_{\tg{i}}}\nonumber
\end{eqnarray}
$\forall i \in \{1,\dots,m\}$. A physicists proof of the theorem follows.
\begin{theorem}
        Let ${\bf M}_{\tb{1:N}_{\tg{m}}}$ be a real tridiagonal $N_{\tb{m}}\times N_{\tb{m}}$ matrix, with entries \\
        $[{\bf M}]_{\tb{ij}} = \delta_{\tb{i-1,j}} \xi^{-1}_{\tb{f(j)}}+ \delta_{\tb{i+1,j}}\xi_{\tb{f(j)}}^{-1}-2\delta_{\tb{ij}}\xi_{\tb{f(j)}}^{-1}$\\
        $\forall i,j\in\{1,\ldots,N_{\tb{m}}\}$. The matrix is partitioned into $m$ segments, where $N_{\tb{m}}=\sum_{l=1}^m n_{\tb{l}}$, \textcolor{black}{with $n_{\tb{l}}\ge 2$ $\forall$ $l\in\{1,..m\}$. For any $p\in\{1,..N_{\tb{m}}\}$}, the $p^{th}$ eigenvalue $\lambda_{\tb{p}}$ is:
\begin{equation}
        \lambda_{\tb{p}}=\frac{4}{\xi_{\tb{f(p)}}}\sin^2\left(\frac{p\pi - \phi_{\tb{m}}}{2(1+N_{\tb{m}})}\right)\nonumber
\end{equation}
        where $\phi_{\tb{m}}=-(N_{\tb{m}}-n_{\tb{m}})k+\tan^{-1}(\frac{v_{\tb{m}} \sin(k_{\tb{p}})}{u_{\tb{m}}})$, $k_{\tb{p}}=\frac{p\pi - \phi_{\tb{m}}(k_{\tb{p}})}{(1+N_{\tb{m}})}$ and $u_{\tb{m}}$ and $v_{\tb{m}}$ are defined by the following products of matrices,
\begin{equation}
        \label{qmpm1}
\begin{bmatrix}
    u_{\tb{m}}        \\
    v_{\tb{m}}        
\end{bmatrix}
= 
        \left(\prod_{j=m}^{2} \begin{bmatrix}
        F_{\tb{n}_{\tg{j-1}}}(k)  &  Q_{\tb{n}_{\tg{j-1}}}(k)      \\
                w_{\tb{j-1}} \mathcal{U}_{\tb{n}_{\tg{j-1}}\tb{-1}}(k)  & w_{\tb{j-1}} \mathcal{T}_{\tb{n}_{\tg{j-1}}}(k)      
\end{bmatrix} 
        \right)
\begin{bmatrix}
        1        \\
        0        
\end{bmatrix}
\end{equation}
\end{theorem}

\textcolor{black}{Based on the cases $m=1$ and $m=2$, we can make an induction hypothesis for the form of the characteristic polynomial for the matrix ${\bf M}_{1:N{\tb{s}}}$, for any $m=s$ to be}:
\begin{equation}
	\label{fullsec}
	\tpsi_{1:N_{\tb{s}} }= \frac{\sin[(N_{\tb{s}}+1) k + \phi_{\tb{s}}]}{\sin(k)}
\end{equation}
We can expand $\tpsi_{\tb{1:N}_{\tg{s+1}}}$ using Laplace's expansion between submatrices $\tpsi_{\tb{1:N}_{\tg s}}$ and 
$\tpsi_{{\tb{N}_{\tg{s}}}+\tb{1}:\tb{N}_{\tg{s+1}}}$ as follows:
\begin{equation}
        \label{laplaceexp}
        \tpsi_{\tb{1:N}_{\tg{s+1}}}=\tpsi_{\tb{1:N}_{\tg{s}}}\mathcal{U}_{\tb{n}_{\tg{s+1}}} -w_{\tb{s}} \tpsi_{\tb{1:N}_{\tg{s}}}\mathcal{U}_{\tb{n}_{\tg{s+1}}\tb{-1}}
\end{equation}
Using identity (\ref{id1}), we have:
\begin{equation}
        \label{fullsec2}
        \tpsi_{\tb{1:N}_{\tg{s+1}}} = \frac{\sin[(N_{\tb{s+1}}+1) k + \phi_{\tb{s+1}}]}{\sin(k)}
\end{equation}
with,
\begin{eqnarray}
        \label{fullphi}
	\phi_{\tb{s+1}} &=& -(N_{\tb{s+1}}-n_{\tb{s+1}})k\\\nonumber 
	          &+&\tan^{-1}\left[\frac{w_{\tb{s}} \tpsi_{\tb{1:N}_{\tg{s}}}\sin(k)}{\tpsi_{\tb{1:N}_{\tg{s+1}} }-w_{\tb{s}} \cos(k)\tpsi_{\tb{1:N}_{\tg{s-1}}}}\right]
\end{eqnarray}
Hence, by induction identity (\ref{fullsec}) is true. Then for any $s=m$,
the roots of the polynomial $\tpsi_{\tb{1:N}_{\tg{m}}}=0$ imply:
\begin{equation}
	\label{roots}
	(N_{\tb{m}}+1)k_{\tb{p}} + \phi_{\tb{m}}(k_{\tb{p}}) = p\pi
\end{equation}
From (\ref{laplaceexp}) we see that $\tpsi_{\tb{1:N}_{\tg{m}}}$ can similarly be expanded using polynomials $\mathcal{U}_{\tb{n}_{\tg{m}}}$, it follows that $\tpsi_{\tb{1:N}_{\tg{m}}}$ is a polynomial in $2cos(k)$.
Again, identifying $2-\xi_{\tb{f(p)}}\lambda=2\cos(k_{\tb{p}})$, we have: 
\begin{equation}
	\lambda_{\tb{p}}=\frac{4}{\xi_{\tb{f(p)}}}\sin^2\left(\frac{p\pi - \phi_{\tb{m}}}{2(N_{\tb{m}}+1)}\right)
	\label{lambdap}
\end{equation}
$\forall p \in\{1,\ldots, N_{\tg{m}}\}$. $k_{\tb{p}}$ needs to be numerically determined using (\ref{roots}). 
To obtain $\phi_{\tb{m}}$, we need to solve the recurrence relation (\ref{fullphi}). Substituting (\ref{id1}) in  (\ref{laplaceexp}) with $s=m$ :
\begin{eqnarray}
	\label{psinm}
	\tpsi_{\tb{1:N}_{\tg{m}}}&=&\mathcal{U}_{\tb{n}_{\tg{m}}}\left(\tpsi_{\tb{1:N}_{\tg{m-1}}} -w_{\tb{m-1}}\cos(k)\tpsi_{\tb{1:N}_{\tg{m-1}}}\right)\\\nonumber
			&+&w_{\tb{m-1}}\tpsi_{\tb{1: N}_{\tg{m-1}}\tb{-1}} \mathcal{T}_{\tb{n}_{\tg{m}}\tb{+1}}\nonumber
\end{eqnarray}
We propose an ansatz that allows us to express (\ref{psinm}) in terms of certain polynomials $u_{(\tb{m})}$ and $v_{(\tb{m})}$:
\begin{eqnarray}
	\label{psipmqm}
	\tpsi_{\tb{1:N}_{\tg{m}}}&=&u_{(\tb{m})}\mathcal{U}_{\tb{n}_{\tg{m}}} + v_{(\tb{m})}\mathcal{T}_{\tb{n}_{\tg{m}}\tb{+1}}\\\nonumber
	\tpsi_{\tb{1:N}_{\tg{m-1}}}&=&u_{(\tb{m-1})}\mathcal{U}_{\tb{n}_{\tg{m-1}}} + v_{(\tb{m-1})}\mathcal{T}_{\tb{n}_{\tg{m-1}}\tb{+1}}\\\nonumber
	\tpsi_{\tb{1:N}_{\tg{m-1}} \tb{-1}}&=&u_{(\tb{m-1})}\mathcal{U}_{\tb{n}_{\tg{m-1}}\tb{-1}} + v_{(\tb{m-1})}\mathcal{T}_{\tb{n}_{\tg{m-1}}}\nonumber
\end{eqnarray}

Substituting (\ref{psipmqm}) in (\ref{psinm}) we have the following relation:
\begin{eqnarray}
	u_{(\tb{m})}&=& u_{(\tb{m-1})} F_{\tb{n}_{\tg{m-1}}}(k) + v_{(\tb{m-1})}Q_{\tb{n}_{\tg{m-1}}}(k)\\\nonumber
  v_{(\tb{m})}&=& u_{(\tb{m-1})}w_{\tb{m-1}} \mathcal{U}_{\tb{n}_{\tg{m}}\tb{-1}} + v_{(\tb{m-1})}w_{\tb{m-1}}\mathcal{T}_{\tb{n}_{\tg{m}}\tb{-1}}\nonumber
\end{eqnarray}	
This allows us to write the recurrence relation in the following form,
\begin{equation}
\begin{bmatrix}
    u_{\tb{m}}        \\
    v_{\tb{m}}        
\end{bmatrix}
= 
\begin{bmatrix}
	F_{\tb{n}_{m-1}}(k)  &  Q_{\tb{n}_{\tg{m-1}}}(k)      \\
	w_{\tb{m-1}} \mathcal{U}_{\tb{n}_{\tg{m-1}}\tb{-1}}(k)  & w_{\tb{m-1}} \mathcal{T}_{\tb{n}_{\tg{m-1}}}(k)      
\end{bmatrix} 
\begin{bmatrix}
	u_{(\tb{m-1})}        \\
        v_{(\tb{m-1})}        
\end{bmatrix}
	\label{justpmqm}
\end{equation}

\textcolor{black}{Starting with $u_{\tb{1}}=1$ and $v_{\tb{1}}=0$ in (\ref{justpmqm}) to express $u_{\tb{m}}$ and $v_{\tb{m}}$ proves (\ref{qmpm1})}. The formula for the eigenvalues (\ref{lambdap}) bears resemblance to the RBP eigenvalues~\cite{Verdier1966,Doi}, but with additional phase shift $\phi_m$, which is expressed using Chebyshev polynomials. The eigenvalues are exact, with a caveat that numerical solving of $k_p$ with (\ref{roots}) is necessary. Using (\ref{singlemode}) and (\ref{lambdap}) allows one to write the $p^{th}$ mode dynamics for rRBC in an uncoupled manner:
\begin{equation}
	\dot{X}_{\tb{p}} = \frac{X_{\tb{p}}(t)}{\lambda^{-1}_{\tb{p}}} + \eta(t)
\end{equation}	
Here, $\lambda^{-1}_{\tb{p}}$ is the relaxation time and $X_{\tb{p}} (t)$ is $p^{th}$ eigenvector of {\bf M}, $\forall p\in \{1,\ldots,N_{\tb{m}}\}$ . The eigenvectors are easily obtained from the eigenvalues~\cite{Tao}.

{\bf Spectral distribution:}
To calculate the spectral distribution $P(\lambda)$ of eigenvalues $\lambda(k,\tphi_m(n),\xi)$, where $\tphi_m=\tan^{-1}(\frac{v_{(m)}\sin(k)}{u_{(m)}})$, we need the following distributions: (i) $h(\tphi_{\tb{m}})$: distribution of the phase $\tphi_{\tb{m}}$, (ii) $\rho(k)$: distribution of the roots $k$, (iii) $P_{\tb{\xi}}(\xi)$: distribution of $\xi$, (iv) $P_{\tb{n}}(n)$: The distribution of  bead-segment size $n$. 	

\textcolor{black}{We startby calculating $h(\tphi_{\tb{m}})$ for arbitrary $P_{\tb{\xi}}(\xi)$ and $P_{\tb{n}}(n)$. We can write equation (\ref{justpmqm}) as a stochastic reccurence relation}:
\begin{equation}
	\tan(\tphi_{\tb{m})}=\sin(k)\frac{w_{\tb{m-1}}\mathcal{U}_{\tb{n}_{\tg{m-1}}-1}+w_{\tb{m-1}}\mathcal{T}_{\tb{n}_{\tg{m-1}}}\tan(\tphi_{\tb{m-1}})}{F_{\tb{n}_{\tg{m-1}}}(k)+\tan(\tphi_{\tb{m-1}})Q_{\tb{n}_{\tg{m-1}}}(k)}
	\label{recctan}
\end{equation}
\textcolor{black}{We introduce a distribution of $\tan(\tphi_{\tb{m}})$: $g(\tan(\tphi_{\tb{m}}))$.} The random term $w$ in (\ref{recctan}) is a ratio of two random variables, $\xi$, and its distribution is $P_{\tb{w}}(w)$ can be obtained from $P_{\tb{\xi}}(\xi)$ using a Mellin transform~\cite{Dale}.
\textcolor{black}{One way to interpret equation (\ref{justpmqm}) is as an evolution of a random walk in the $u$-$v$ space, under the action of a $2\times 2$ random matrix}~\cite{Wu}.
Since $u_{\tb{m}}$ and $v_{\tb{m}}$ are the components of the same vector, they have the same Lyapunov exponent \cite{ProdRand}, hence for a large $m$'s and realization of $w$'s one would expect $g(\tan(\tphi_{\tb{m}}))$ to converge to a {\it limiting distribution}~\cite{Wu}.
 \textcolor{black}{On multiplication of the random matrix the limiting distribution $g(\tan(\tphi_{\tb{m}}))$ transforms to itself and we can write this as an integral equation, with} $r=\tan(\tphi_{\tb{m}})$, $\tilde{r}=\tan(\tphi_{\tb{m-1}})$ as:
\begin{eqnarray}
        g(r)&=&\int_{-\infty}^{\infty}\int_{0}^{\infty}\int_{0}^{\infty}g(\tilde{r})\delta\left(r-\gamma(\tilde{r},w,n)\right)\\\nonumber
            &&P_{\tb{n}}(n)P_{\tb{w}}(w)\:dw\:dn\:d\tilde{r}
        \label{gr}
\end{eqnarray}
Here,
\begin{equation}
        \gamma(\tilde{r},w,n)=\sin(k)\frac{w\mathcal{U}_{\tb{n-1}}+w\mathcal{T}_{\tb{n}}\tilde{r}}{F_{\tb{n}}+\tilde{r} Q_{\tb{n}}}
\end{equation}
Integrating over $w$ first and then $n$ gives:
\begin{eqnarray}
                \label{gr2}
        g(r)&=&\int_{-\infty}^{\infty}g(\tilde{r})\left<\frac{P_w\left( w^*\right)}{\gamma'(\tilde{r},w^*,n)}\right>_n \:d\tilde{r}\\\nonumber
		w* &=&\frac{r\mathcal{U}_{\tb{n}}+r\tilde{r}\mathcal{T}_{\tb{n+1}}}{(\sin(k)+r\cos(k))(\mathcal{U}_{\tb{n-1}}+\tilde{r}\mathcal{T}_{\tb{n}})}\\\nonumber\\\nonumber
\end{eqnarray}
Here, $\left<.\right>_n$ denotes the average over $n$, \textcolor{black}{same convention is used for other variables}.
Here, $\gamma'(r,w,n)=d\gamma(r,w,n)/dw$ is the Jacobian when we transform from $\tilde{r}$ to $w$ variables. Furstenberg's theory~\cite{Furstenberg, Wu} guarantees the existence of a unique solution to the integral equation (\ref{gr2}). 

We first calculate $g(r)$ from (\ref{gr2}) (See Supplememtary Material)~\cite{supp}.
Since, $h(\phi)d\phi=g(r)dr$, we have:
\begin{equation}
        h(\phi) = g(r)\left(1+r^2\right)    
\end{equation}
From (\ref{roots}) the density of the roots $\rho(k)=\frac{d p}{d k_{\tb{p}}}$ is:
\begin{equation}
	\rho(k)=(N_{\tb{m}}+1+\tphi_{\tb{m}}^{'})/\pi
	\label{rootdensity}
\end{equation}
Here, $\tphi_m^{'}=d\tphi_{\tb{m}}/dk$. 
We choose a simplifying distribution $P_{\tb{n}}(n)\propto \exp\left[-\alpha (n-n_o)^2\right]$, $\alpha$ is very large.
We can now calculate the  distribution of eigenvalues  $P(\lambda)$:
\begin{eqnarray}\nonumber
	P(\lambda)& =& \int_{0}^{\infty}\int_{0}^{\pi}\int_{0}^{\pi} \rho(k)\delta(\lambda-\lambda(k,\xi,\phi(n))) \\
	& & dk\: P_{\tb{\xi}}(\xi)\: d\xi\: h(\phi)\: d\phi\: P_{\tb{n}}(n)\:dn\\\nonumber
	P(\lambda)&\propto& \left[ \frac{1}{\pi} \int_{0^{+}}^{(\frac{4}{\lambda})^{-}}\xi \frac{P_{\tb{\xi}}(\xi)\: d\xi}{\sqrt{(\frac{\lambda\xi}{4})(1-\frac{\lambda\xi}{4})}}\right]\\\nonumber
		  &+&\left[ \frac{1}{\pi N_{\tb{m}}}\int_{0^{+}}^{(\frac{4}{\lambda})^{-}} \frac{\xi \left<\tphi^{'}(\sin^{-1}(\sqrt\frac{\lambda\xi}{4})\right>_{\phi} P_{\tb{\xi}}(\xi)}{\sqrt{(\frac{\lambda\xi}{4})(1-\frac{\lambda\xi}{4})}}\:d\xi \right] \\
	\label{limspectra}		  
\end{eqnarray}
\begin{figure}[t]
\includegraphics[width=3.0in]{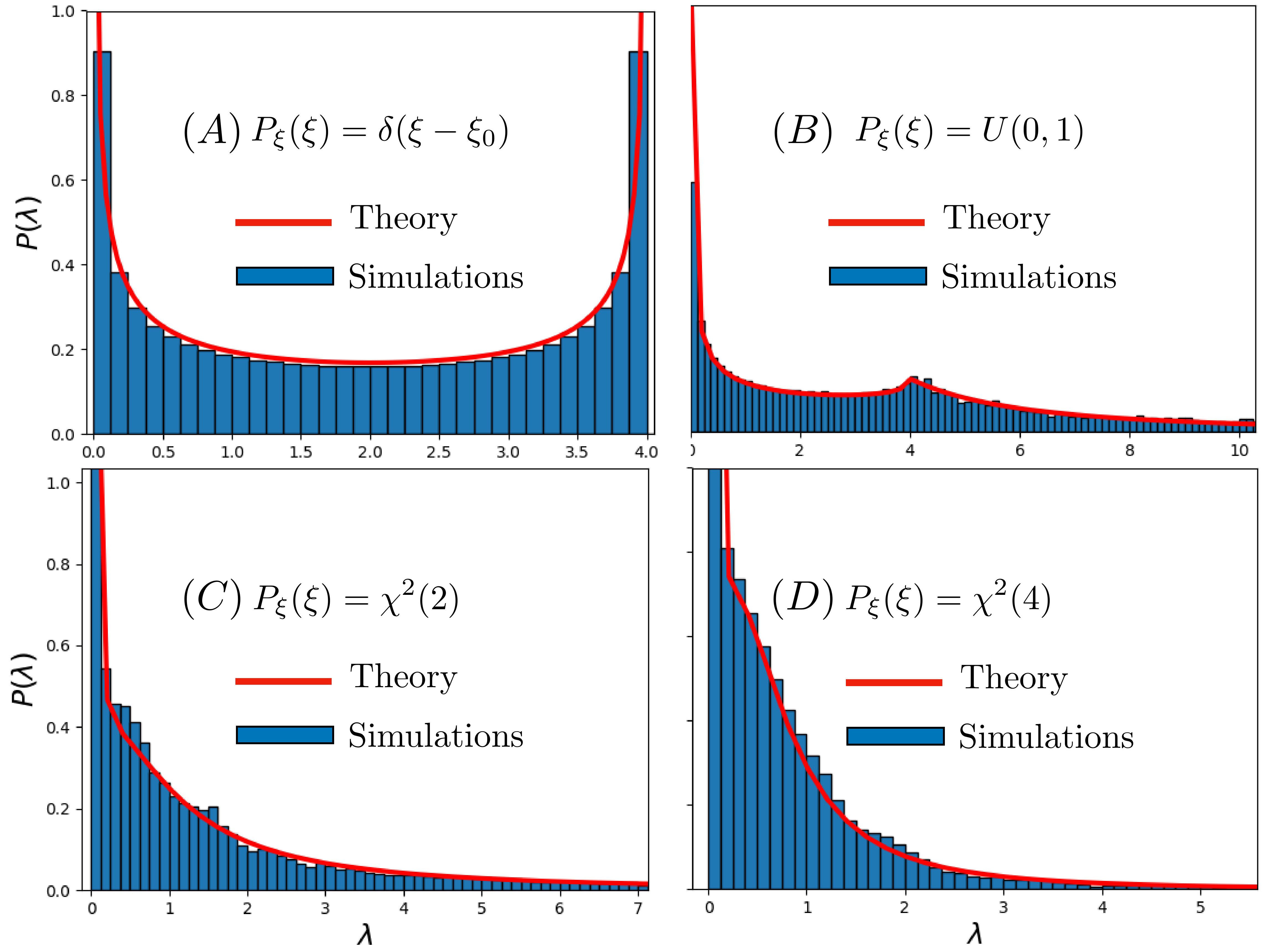}
	\caption{  $P(\lambda)$ for distribution (i) $P(\xi)=\delta(\xi-\xi_o)$ (ii) $P(\xi)$ is uniform (iii) $P(\xi)$ is $\chi^2(2)$, (iv)$P(\xi)$ is $\chi^2(4)$. Histograms obtained from eigenvalues of $10000\times10000$ matrix, $n_o=100$ segments. Theoretical curves are in red. }  
\label{fig1}
\end{figure}
\textcolor{black}{The second term in equation (\ref{limspectra}) is $\mathcal{O}(\frac{1}{N_{\tb{m}}})$. Although the $n_i$'s need only be $\ge2$ in (\ref{qmpm1}), when $N_m$ is large, the effect boundary conditions between segments which comes through $h(\phi_m)$ is inconsequential. This highly simplifies the expression of $P(\lambda)$. In the limit of $P_{\tb{\xi}}(\xi)\rightarrow \delta(\xi-\xi_o)$, the spectral distribution of tridiagonal Toeplitz matrix is recovered from (\ref{limspectra}). Fig.(\ref{fig1}) shows the spectral distribution obtained using equation (\ref{limspectra}) for three cases: (i) $P_{const}(\lambda)$ for $P_{\xi}(\xi)=\delta(\xi-\xi_o)$, (ii) $P_{uni}(\lambda)$ for $P_{\xi}(\xi)=U(0,1)$: uniform distribution for $(0<\xi<1)$ and $0$ elsewhere, and (iii) $P_{\chi^2(d)}(\lambda)$ for $P_{\xi}(\xi)=\chi^2(d)$ distribution with degree of freedom $d$}.
\begin{equation}
	P_{const}(\lambda)\propto\frac{\xi_o}{\sqrt{\frac{\xi_o\lambda}{4}(1-\frac{\xi_o\lambda}{4}})}\\
\end{equation}
\begin{eqnarray}
	P_{uni}(\lambda)&\propto&
	\begin{cases}
		
		\frac{16\sin^{1}(\frac{\sqrt{\lambda}}{2})}{\lambda^2}-\frac{4\sqrt{\lambda(4-\lambda)}}{\lambda^2},& \text{if } ~0\leq \lambda \leq4 \\
		\frac{8\pi}{\lambda^2},& \text{if } ~\lambda > 4 \\
    \end{cases}
\end{eqnarray}	
\begin{equation}
	P_{\chi^2(d)}(\lambda)\propto\frac{1}{\lambda^{d/2+1}} \int_{0^+}^{1^{-}}\frac{y^{\frac{d}{2}} e^{-2y/\lambda}}{\sqrt{y(1-y)}}dy
\end{equation}	
\textcolor{black}{ Fig. 1(a) with (b-d) captures the effect of disorder on the mode relaxation times ($\lambda^{-1}$), one sees that the concentration eigenvalues at $\lambda=4$ decrease and spread beyond $\lambda>4$. Most studies of polymers in random media has been done using replica theory~\cite{Yadin1,Yadin2,Muthu}, to my knowledge this is the first exact results of polymers in random media without replica theory, showing disorder induced relaxation time spread. To see this effect on dynamical quantities like MSD, one needs to obtain $\vec{R}$ which is weighted summation over $\vec{X}_p$ and the weights depend on $\lambda_p$~\cite{Doi}. How the summation of this series effects the scaling of MSD with time $t$ and localization transition~\cite{Vilgis2000} will be reported in a subsequent publication.}

\textcolor{black}{We have calculated exact closed form eigenvalues for rsqt-T, this has enabled the diagonalization of the Rouse modes for the rRBC. Furstenburgs theory enables calculation of analytical spectral distribution. We have shown that the boundary conditions between the random segments do not effect the spectral distribution for large polymers. Implications of this study in live cells nucleosomes is under investigation.}

I wish to thank Deepak Dhar for his comments on the manuscript and making me aware of some of the important related literature.  I thank  Vinod Krishna, Bhaswati Bhattacharyya, Marimuthu Krishnan and Masaki Sasai for their comments. I thank Masaki Sasai and Marimuthu Krishnan for their support. I acknowledge financial support from CREST Grant JPMJCR15G2 of Japan Science and Technology Agency and the KAKENHI Grants, 20H05530, and 21H00248 of Japan  Society for the Promotion of  Science. This research was supported in part by the International Centre for Theoretical Sciences (ICTS) for the online meeting - Celebrating the Science of Giorgio Parisi (code: ICTS/CSGP2021/12).


\begin{thebibliography}{999}
        \bibitem{Gautam}G.I. Menon, P. Ray,``The physics of disordered systems", TRiPS, {\bf7}, Springer Publisher, (2012).
        \bibitem{Bovier} Anton Bovier, "Statistical Mechanics of disordered systems: Mathematical perspective", Cambridge University Press, (2006).
        \bibitem{MLMehta}M. L. Mehta, "Random Matrices",3rd Edition, Elsevier press, (2004).
        \bibitem{Vivo}G. Livan, M. Novaes, P. Vivo, Introduction to random matrices: theory and practice, Springer (2018).
        \bibitem{Potters}M. Potters, J.-P. Bouchaud, “A First Course in Random Matrix Theory, Cambridge University Press (2020).
        \bibitem{PJForrester2003}PJ Forrester, NC Snaith, JJM Verbaarschot, J. Phys. A: Mathematical and General. {\bf 36}, R1, (2003).
        \bibitem{PJForrester1}J. Forrester,J. Math. Phys. 62, 103302, (2021).
        \bibitem{Wigner1955}E. Wigner, Annals of Mathematics. {\bf 62} (3): 548–564 (1955); SIAM Rev. {\bf 9}, 1, (1967).
        \bibitem{Dyson1962}F.J. Dyson, J. Math. Phys. {\bf 3}, 140–156, (1962).
        \bibitem{Aljadeff} Aljadeff, M. Stern, and T. Sharpee, Phys. Rev. Lett. {\bf 114}, 088101 (2015).
        \bibitem{Coolen}  A. C. Coolen, P. Sollich, and R. Kühn, Theory of Neural Information Processing Systems (Oxford University Press, Oxford, UK, 2005).
        \bibitem{Kuczala}A. Kuczala and T. O. Sharpee, Phys. Rev. E {\bf 94}, 050101(R) (2016).
        \bibitem{Rajan} K. Rajan and L. F. Abbott, Phys. Rev. Lett. {\bf 97}, 188104 (2006).
        \bibitem{Louart}C. Louart, Z. Liao, and R. Couillet, Ann. Appl. Probab. {\bf 28}, 1190 (2018).
(2018).
        \bibitem{Sachdev} S. Sachdev and J. Ye, Phys. Rev. Lett. {\bf 70}, 3339 (1993).
        \bibitem{Mezard}M. Mezard, G. Parisi, and M. Virasoro, ``Spin Glass Theory and Beyond: An Introduction to the Replica Method and Its
Applications" , World Scientific Publishing Company London, {\bf 9}, (1987).
        \bibitem{Bray}A. J. Bray and M. A. Moore,  J. Phys. C {\bf 12}, L441 (1979).
        \bibitem{Kosterlitz}. M. Kosterlitz, D. J. Thouless, and R. C. Jones, Phys. Rev. Lett. {\bf 36}, 1217 (1976).
        \bibitem{RMMay}R. M. May, Nature(London) {\bf 238}, 413 (1972).
        \bibitem{Barabas}G. Barabás, M. J. Michalska-Smith, and S. Allesina, Nat. Ecol. Evol. {\bf 1}, 1870 (2017)
        \bibitem{Sarnak1996}Z. Rudnick and P. Sarnak, Duke Math. J. {\bf 81},269, (1996).
        \bibitem{Keating1995} E.B. Bogomolny and J.P. Keating,  Nonlinearity {\bf 8},1115, (1995).
        \bibitem{Katz1999}N.M. Katz and P. Sarnak, Random matrices, AMS, Providence, Rhode Island, (1999).
        \bibitem{Bohigas} O. Bohigas, M.J. Giannoni and C. Schmit,  Phys. Rev. Lett. {\bf 52} 1 (1984).
        \bibitem{PJForresterBook}P.J. Forrester, Log-gases and random matrices, Princeton University Press, (2010).
        \bibitem{Albrecht}J.T. Albrecht, CY. P. Chan, A. Edelman, Found Comput Math 9, 461–483 (2009).
        \bibitem{Baron2022} J. W. Baron, T. J. Jewell, C. Ryder , T. Galla, Phys. Rev. Lett. {\bf 128},  120601 (2022).
         \bibitem{Tao} Peter B. Denton, Stephen J. Parke, Terence Tao, Xining Zhang Bull. Amer. Math. Soc. {\bf 59} 31-58, (2022).
        \bibitem{Dean}P. Dean, Proc Phys Soc {\bf 84}, 727 (1964)
        \bibitem{Hori} J. Hori, "Spectral properties of disordered chains and lattices", Pergamon press, (1968).
        \bibitem{Matsuda}H. Matsuda and K. Ishi, Prog. Theor Phys Suppl {\bf 45} 56-86 (1970).
        \bibitem{Dyson}F. J. Dyson, Phys. Rev. 92, 1331 (1953)
        \bibitem{Toep1911}O. Toeplitz, Rend. Cire. Mat. Palermo, {\bf 32},191–192, (1911).
        \bibitem{Gray}R. M. Gray, Toeplitz and Circulant Matrices: A review (Now Pub, 2006).
        \bibitem{Onsager}B. Kaufman, L. Onsager, Phys. Rev. 76 (1949), no. 8, 1244–1252.
        \bibitem{EBogomolny}E. Bogomolny, Phys. Rev. E {\bf 102}, 040101(R), (2020).
        \bibitem{Hartwig}M. Fisher and Hartwig, Advances in Chemical Physics 15, 333 (1968).
        \bibitem{Qin}B.-Q.Jin and V.E.Korepin, Journal of Statistical Physics 116, 79 (2004).
        \bibitem{Ivanov}D. A. Ivanov and A. G. Abanov, Journal of Physics A: Mathematical and Theoretical 46, 375005 (2013).
        \bibitem{Keat} J. Keating, F. Mezzadri, Commun. Math. Phys. {\bf 252}, 543D579 (2004).
        \bibitem{Dunlop} D.H. Dunlap, H-L. Wu, P. W. Philips, Phys. Rev. Lett. {\bf 65}, 88, (1990).
        \bibitem{EpigenRev}S. Kim, B.K. Kaang, Exp Mol Med 49, e281 (2017).
        \bibitem{Virk2020}R. K. A. Virk, W. Wu, L. M. Almassalha, G. M. Bauer, Y. Li, D. VanDerway, J. Frederick, D. Zhang, A. Eshein, H. K. Roy, I. Szleifer, V. Backman, Sci. Adv. 6, eaax6232 (2020).
        \bibitem{AshPNAS} S.S. Ashwin, T. Nozaki, K. Maeshima, M. Sasai. Proc Natl Acad Sci USA. 116, 19939-19944, (2019).
        \bibitem{AshRev} S.S Ashwin, K. Maeshima, M. Sasai, Biophysical Reviews. 12, 461–468, (2020).
        \bibitem{AmitaiRev}A. Amitai, D. Holcman, Physics Reports {\bf 678}, 1-83 (2017).
        \bibitem{Itoh2021} Itoh, Y., Woods, E.J., Minami, K., Maeshima, K., Collepardo-Guevara, R,Current Opinion in Structural Biology. 71, 123–135 (2021).
        \bibitem{Maeshima2020} Maeshima, K., Tamura, S., Hansen, J.C., Itoh, Y.,Current Opinion in Cell Biology. {\bf 64}, 77-89,(2020).
        \bibitem{Courchaine} E.M. Courchaine, A. Lu, K.M. Neugebauer, EMBO J.l, {\bf 35} (15), 1603-1612, (2016).
        \bibitem{Maeshima2019}Prieto, E.I., Maeshima, K.,Essays in Biochemistry {\bf 63}, 133-145 (2019)
        \bibitem{MaeshimaII2019} K. Maeshima, S. Ide, M. Babokhov, Current Opinion in Cell Biology {\bf 58}, 95–104 (2019).
        \bibitem{Sasai2016}K. Maeshima, S. Ide, K. Hibino, M. Sasai, Current Opinion in Genetics and Development {\bf 37}, 36–45 (2016).
        \bibitem{Verdier1966} P. H. Verdier, J. Chem. Phys., {\bf  45}, 2118-212 (1966).
        \bibitem{Doi} M. Doi and S.F. Edwards, The theory of polymer dynamics, Oxford: Clarendon Press (1986).
        \bibitem{HolcmanPRE} O. Shukron and D. Holcman, Phys. Rev. E {\bf 96}, 012503, (2017).
        \bibitem{Socol} M. Socol, R. Wang, D. Jost, P. Carrivain, C. Vaillant, E.L. Cam, V. Dahirel, C. Normand, K. Bystricky, J-M. Victor, O. Gadal, A. Bancaud, Nucleic Acids Research, {\bf 47}, 12, 6195–6207, (2019).
        \bibitem{Nozaki2017}T. Nozaki, R. Imai, M. Tanbo, R. Nagashima, S. Tamura, T. Tani, Y. Joti,M. Tomita, K. Hibino, M.T. Kanemaki, K.S. Wendt,Y. Okada, T. Nagai, K. Maeshima.  Mol Cell.{\bf 67},2,282-293.e7, (2017).
        \bibitem{Winkler}R. G. Winkler and G. Gompper, J. Chem. Phys. {\bf 153}, 040901, (2020).
        \bibitem{Rabin}D. Osmanovic, Y. Rabin, Soft Matter, {\bf 13}, 963-968, (2017).
        \bibitem{Chaikin} P.M. Chaikin and T.C. Lubensky, Principles of condensed matter physics, Cambridge University Press, Cambridge, (1995).
        \bibitem{Grudsky}S.M Grudsky,E.A. Maximenko, A. Soto-González, Chapter in the book: Karapetyants, Operator Theory and Harmonic Analysis. OTHA 2020. Springer Proceedings in Mathematics \& Statistics, {\bf 357} Springer, Cham.
        \bibitem{Mason}J.C. Mason, and D. C. Handscomb. Chebyshev polynomials. CRC press, (2002).
        \bibitem{Kulkarni} D. Kulkarni, D. Schmidt,  Sze-Kai Tsui Linear Algebra and its Applications, {\bf 297}, 63-80, (1999).
        \bibitem{Banchi}L. Banchi and R. Vaila, J. Math. Phys. {\bf 54}, 043501, (2013).
        \bibitem{Dale} Melvin Dale, "The Algebra of Random variables", Wiley, New York, (1979).
        \bibitem{Wu} B.M McCoy and T.T. Wu,  Phys. Rev. {\bf 176}, 631, (1968).
        \bibitem{ProdRand} A Cristani, G. Paladin, A. Vulpiani, {\it Products of Random Matrices}, Springer-Verlag, (1993).
        \bibitem{Furstenberg} H. Furstenberg, Trans. Am. Math. Soc. 108, 377 (1963).
        \bibitem{supp} Supplementary Material.
	\bibitem{Yadin1}Y. Shiferaw and Y. Y Goldschmidt,Journal of Physics A: Mathematical and General, {\bf 33}, 4461, (2000).
	\bibitem{Yadin2}Y. Y Goldschmidt, Phys. Rev. E, {\bf 61}, 1729, (2000).
	\bibitem{Muthu}SF Edwards and M Muthukumar, Journal Chem. Phys, {\bf 89},2435–2441, (1988).	
        \bibitem{Vilgis2000} T.A. Vilgis, Physics Reports 336, 167,254, (2002).
\end{thebibliography}
\end{document}